\documentclass[a4paper,fleqn,usenatbib,useAMS]{jaa}
\usepackage[english]{babel}

\def\k{{\bf k}}

\def\u{\vec{U}}

\def\mnras{MNRAS} 
\def\jcap{JCAP}
 
\def\aap{A \& A}
\def\apj{Ap.J}

\def\u{{\bf U}}

\def\lsim{~\rlap{$<$}{\lower1.0ex\hbox{$\sim$}}}

\def\gsim{~\rlap{$>$}{\lower 1.0ex\hbox{$\sim$}}}

\usepackage{graphics} 
\usepackage{graphicx} 
\usepackage{color}
\usepackage{amsmath} 
\usepackage{psfrag} 
\usepackage{epsfig}

\begin{document}

\title[Binned HI power spectrum for OWFA] {Fisher matrix based 
predictions for measuring the $z=3.35$  binned 21-cm power spectrum  
using the Ooty Wide Field Array (OWFA)}
\author[A.K. Sarkar, S. Bharadwaj and S. S. Ali] {Anjan Kumar
  Sarkar$^{1}$\thanks{Email:anjan@cts.iitkgp.ernet.in}, Somnath
  Bharadwaj$^{2,1}$\thanks{Email:somnath@phy.iitkgp.ernet.in} and
  Sk. Saiyad Ali$^{3}$\thanks{Email:saiyad@phys.jdvu.ac.in}\\ $^{1}$
  Centre for Theoretical Studies, IIT Kharagpur, 721302, India
  \\ $^{2}$ Department of Physics, IIT Kharagpur, 721302, India
  \\ $^{3}$ Department of Physics, Jadavpur University, Kolkata
  700032, India} \date {} \maketitle

\begin{abstract}
\noindent We use the Fisher matrix formalism to predict the prospects
of measuring the redshifted 21-cm power spectrum in different $k$-bins
using observations with the upcoming Ooty Wide Field Array (OWFA)
which will operate at $326.5 {\rm MHZ}$. This corresponds to neutral
hydrogen (HI) at $z=3.35$, and a measurement of the 21-cm power
spectrum provides an unique method to probe the large-scale structures
at this redshift. Our analysis indicates that a $5 \sigma$ detection
of the binned power spectrum is possible in the $k$ range $0.05 \leq k
\leq 0.3 \, {\rm Mpc}^{-1}$ with $1,000$ hours of observation.  We
find that the Signal-to-Noise ratio (${\rm SNR}$) peaks in the $k$
range $0.1- 0.2\, {\rm Mpc}^{-1}$ where a $10 \sigma$ detection is
possible with $2,000$ hours of observations. Our analysis also
indicates that it is not very advantageous to observe much beyond
$1,000$ hours in a single field of view as the ${\rm SNR}$ increases
rather slowly beyond this in many of the small $k$-bins. The entire
analysis reported here assumes that the foregrounds have been
completely removed.
\end{abstract}


\section{Introduction}
The redshifted 21-cm emission from the discrete, unresolved neutral
hydrogen (HI) sources in the post-reionization era ($z < 6$) appears
as a faint diffuse background radiation in low frequency observations
below 1420 MHz. This provides us a useful tool to explore the large
scale structure of the universe in the post-reionization era, using
the fluctuations in the diffuse background radiation to trace the HI
power spectrum (Bharadwaj, Nath \& Sethi 2001; Bharadwaj \& Sethi
2001). In addition to probing the HI power spectrum (Bharadwaj \&
Pandey 2003; Bharadwaj \& Srikant 2004), fluctuations in the diffuse
background radiation also probe of the bispectrum (Ali et al. 2006;
Guha Sarkar \& Hazra 2013). In recent years, a considerable amount of
work has been done to explore the prospects of detecting the 21-cm HI
signal from the post-reionization era (Visbal et al. 2009; Bharadwaj
et al. 2009; Wyithe \& Loeb 2009; Seo et al. 2010; Mao 2012; Ansari et
al. 2012; Bull et al. 2014).

 In the post-reionization era, the bulk of the HI 21-cm emission
 originates from the dense pockets of self-shielded HI regions, which
 are identified as damped Ly$\alpha$ (DLA) systems in quasar
 observations.  The fluctuations in the HI 21-cm emission which are in
 general quantified through HI power spectrum, is expected to trace
 the matter power spectrum with a possible bias (Bharadwaj, Nath \&
 Sethi 2001; Bharadwaj \& Sethi 2001). Wyithe \& Loeb (2009) have
 shown that the complications in the HI power spectrum arising due to
 the modulation of the ionizing field are less than 1\%. Bagla,
 Khandai \& Datta (2010) have used semi-numerical simulations to
 predict the HI bias. They have used three different prescriptions to
 assign HI mass to the dark matter haloes and have found the HI bias
 to be scale-independent on large scales ($k \leq 1$ Mpc$^{-1}$). Guha
 Sarkar et al. (2012) and Vilaescusa Navarro et al. (2014) have used
 simillar simulations and their results are also found to be
 consistent with a scale-independent HI bias on large
 scales. Recently, Sarkar, Bharadwaj \& Anathpindika 2016 have used
 semi-numerical simulations to model the HI bias and have provided a
 fitting formula for the HI bias $b_{HI}(k,z)$ in both $k$ and $z$
 across $0.01 \leq k \leq 10$ Mpc$^{-1}$ and $0 \leq z \leq 6$.

The measurement of the HI power spectrum holds the possibility of
constraining the background cosmological model through the Baryon
Acoustic Oscillations (BAO) (Wyithe et al. 2008; Chang et
al. 2008). Seo et al. (2010) have studied the possibility of measuring
BAO using the HI power spectrum with a ground-based radio
telescope. The measurement of the HI power
spectrum can also be used to constrain cosmological parameters
independent of the BAO (Bharadwaj et al. 2009; Visbal, Loeb \& Wyithe
2009). The measurement of the HI power spectrum can also be used to
constrain the neutrino mass (Loeb \& Wyithe 2008). Villaescusa Navarro
et al. (2015) have used hydrodynamical simulations to study the
signatures of massive neutrinos on the HI power spectrum and put
constraints on the neutrino mass. In a recent work, Pal \& Guha Sarkar
(2016) have studied the prospects of measuring the neutrino mass using
the HI 21-cm and the Ly-$\alpha$ forest cross-correlation power
spectrum.

Measurements of the HI power spectrum are sensitive to both the mean
neutral hydrogen fraction $\Omega_{HI}(z)$ and the HI bias
$b_{HI}$. Several measurements have been carried out in past years to
measure the value of the $\Omega_{HI}$ both at low and at high
redshifts. Measurements of the $\Omega_{HI}$ at low redshifts ($z \leq
1$) come from HI galaxy surveys (Zwaan et al. 2005; Martin et
al. 2010; Delhaize et al. 2013), DLAs observations (Rao et al. 2006;
Meiring at al. 2011) and HI stacking (Lah et al. 2007; Rhee at
al. 2013), while measurements of $\Omega_{HI}$ at high redshifts ($1 <
z < 6$) are from DLAs studies (Prochaska \& Wolfe 2009; Norterdaeme et
al. 2012; Zafar et al. 2013). Measurement of the HI power spectrum can
provide astrophysical information about the HI distribution.

Efforts have been made to measure the $\Omega_{HI}$ at redshifts $z <
1$. Chang et al. (2010) and Masui et al. (2013) have studied the
cross-correlation of the HI intensity and the galaxy surveys, while
Switzer et al. (2013) have studied the autocorrelation of HI intensity
to measure the $\Omega_{HI}$. Ghosh et al. (2012) have used the GMRT
observations to place a upper limit on the value of $\Omega_{HI}$ at
$z = 1.33$.

Several low frequency radio interferrometric arrays
(CHIME\footnote{http://chime.phas.ubc.ca/}, Bandura et al. 2014;
BAOBAB\footnote{http://bao.berkeley.edu/}, Pober et al. 2013) are
planned to measure the BAO using the 21-cm signal from $z \leq
2.55$. Shaw et al. (2014) present theoretical estimates for the
sensitivity of CHIME to constrain the line-of-sight and angular scale
of the BAO.  The Giant Meterwave Radio Telescope (GMRT
\footnote{http://gmrt.ncra.tifr.res.in/}, Swarup et al. 1991)
which operates at frequencies corresponding to HI in the redshift
range $0 \leq z \leq 8.5$. The GMRT is currently being upgraded. The
prospects of detecting the HI power spectrum from the post
reionization era for the upgraded GMRT (uGMRT) has been studied in
Chatterjee et al. (2016). The proposed future telescopes SKA1-mid and
SKA1-low \footnote{https://www.skatelescope.org/home/technicaldatainfo/key-documents/},
both hold the prospect of measuring the post-reionization HI power
spectrum at a high level of precision (Guha Sarkar \& Datta 2015; Bull
et al. 2015; Santos et al. 2015).

There is a rich literature on the sensistivity estimates for various
low frequency radio telescopes. Morales et al. (2005) present a
general method to calculate the Epoch of Reionization (EOR) power
spectrum sensitivity for any radio-interferrometric array. Harkar et
al. (2010) have made sensitivity estimates for the Low Frequency Array
(LOFAR{\footnote{http://www.lofar.org/}}, van Haarlem et
al. 2013). Simillar estimates have been made in the context of the
Murchison Wide-field Array
(MWA{\footnote{http://www.mwatelescope.org}}, Bowman et al. 2013;
Tingay et al. 2013). Beardsley et al. (2013) have estimated the
sensitivity of the MWA to the EOR 21-cm power spectrum and McQuinn et
al. (2006) present estimates for cosmological parameter estimation
using MWA5000, a hypothetical extended version of MWA.  Parsons et
al. (2012) have explored redundancy calibration in the context of the
Donald C. Backer Precision Array to Probe the Epoch of Reionization
(PAPER{\footnote{http://astro.berkeley.edu/dbacker/eor}}, Parsons et
al. 2010).  Pober et al. (2014) have studied the constraints
acheiveable with the Hydrogen Epoch of Reionization Array
(HERA{\footnote{http://reionization.org/}}, DeBoer et al. 2016; Neben
et al. 2016) and have found that a very high significance ($\gsim \,
30\sigma$) detection of the reionization power spectrum is possible in
even the most pessimistic scenarios.  Aaron Ewall-Wice et al. (2016)
have used the Fisher matrix formalism to predict the sensitivity with
which it will be possible to constrain reionization and X-ray heating
models with the future HERA and SKA phase I.

Here, we discuss the prospects of measuring the HI power spectrum at
$z \sim 3$, using the upgraded Ooty radio Telescope (ORT). The ORT
consists of a 530 m long and 30 m wide parabolic cyllindrical
reflector, which is placed in the north-south direction on a hill
having the same slope as the latitude (11$^{o}$) of the station
(Swarup et al. 1971; Sarma et al. 1975). It is possible to observe the
same part of the sky through a single rotation of the long axis, which
is aligned with earth's rotation axis. The entire feed system of the
ORT has 1056 dipoles, spaced 0.47 m apart from each other, which are
placed along the focal line of the telescope. The cyllindrical Ooty
Radio Telescope (ORT) is currently being upgraded (Prasad \&
Subrahmanya 2011a, b; Marthi \& Chengalur 2014, Subrahmanya 2017a, b) to function as a
linear radio interferrometric array the Ooty Wide Field Array
(OWFA). The OWFA works at a nominal frequency of $\nu_{0} = 326.5$
MHz, which corresponds to HI radiation from the redshift $z =
3.35$. The OWFA can operate in two independenet interferrometric modes
- Phase I and Phase II. In this work, we have considered Phase II
only. The Phase II has 264 antenna elements, where each antenna
element consists of 4 dipoles. Each antenna has a rectangular aperture
of dimension 1.92 m $\times$ 30 m. The field-of-view of OWFA Phase II is highly
assymetric in dimension, $27.4^{o} \times 1.75^{o}$. The operating
bandwidth for the Phase II is 40 MHz. The field-of-view and the
observing bandwidth of the OWFA Phase II allow to observe the universe over a
real space volume $\sim 0.3 \, {\rm Gpc}^3$.

We now report some recent works related to OWFA. Calibration is an
important issue for OWFA and it has been addressed in Marthi \&
Chengalur (2014). Gehlot \& Bagla (2017) have followed the approach of
Ali \& Bharadwaj (2014) to predict the HI signal expected at
OWFA. Marthi (2017) presents a programmable emulator for simulating
OWFA observations for which foreground modelling and predictions are
presented in Marthi et al. (2017). Chatterjee, Bharadwaj \& Marthi
(2017) present simulations of the HI signal expected at OWFA.

Ali \& Bharadwaj (2014) (hereafter, Paper I) have studied the prospects of detecting the
21-cm signal using OWFA. In this paper, We have also made detailed
foreground predictions for OWFA. In a recent study, Bharadwaj, Sarkar
\& Ali (2015)(hereafter, Paper II) have used the Fisher matrix analysis to make predictions
for hours of observations to measure the HI power spectrum. We showed
that the dominant contribution to the OWFA HI signal is from the
$k$-range $0.02 \leq k \leq 0.2$ Mpc$^{-1}$. It was found that a
$5\sigma$ detection of the HI power spectrum is possible with $\sim
150$ hours of observations using the Phase II. In this study, we have also
explored the possibility of measuring the redshift space distortion
parameter $\beta$. We found that the non-uniform sampling of the
$\mathbf{k}$-modes does not make OWFA suitable for measuring $\beta$.

The predictions for OWFA, mentioned earlier, have all assumed that the
HI power spectrum is related to the $\Lambda$CDM power spectrum with a
scale-independent linear HI bias. All of these studies have focussed
on measuring the amplitude of the HI power spectrum assuming that the
shape of the matter power spectrum (Eisenstein \& Hu 1998) is
precisely known. It is interesting and worthwhile to consider a
situation where both amplitude and the shape of the HI power spectrum
is unknown. There are several astrophysical processes which could in
principle, change the shape of the HI power spectrum without affecting
the matter power spectrum. Further, uncertainties in the background
cosmological model would also be reflected as changes in the observed
HI power spectrum through various effects like redshift space
distortion and Alcock-Paczynski (AP) effect. In this paper, we have
considered the possibility of measuring the HI power spectrum using
OWFA. For this purpose, we have divided the $k$-range into several
bins and employed the Fisher matrix analysis to make predictions for
measuring the HI power spectrum in each of these $k$-bins. Throughout
our analysis, we have used the $\Lambda {\rm CDM}$ cosmology with
PLANCK+WMAP9 best-fit cosmological parameters (Ade at al. 2014).

The paper is structured as follows. In section 2, we present the
theoretical HI model which was used for calculating the signal and
noise covariance. Here we also show the Fisher matrix technique which
was employed for estimating the binned HI power spectrum. In section
3, we use the results from the Fisher matrix analysis to make
predictions for measuring the binned 21-cm power spectrum. We end with
summary and conclusions in Section 4.

\section{Visibility covariance \& Fisher matrix}

OWFA measures visibilities $\mathcal{V}(\mathbf{U}_a, \nu_n)$ at given
baselines $\mathbf{U}_a$ and frequency channel $\nu_{n}$. The baseline
configuration of the OWFA is one-dimensional. It consists of 264
antennas, arranged in a linear array along the length of the
cylinder. Assuming the $\mathbf{x}$-axis to be along the length of the
cylinder, the baselines of the OWFA can be written as follows, i.e,
\begin{equation}
\mathbf{U}_{a} = a \left(\frac{d}{\lambda}
\right)\hat{i} \hspace{2.5cm} (1 \leq a \leq 263)
\end{equation}
where $a$ denotes the baseline number, $d=1.92$ m, is the distance
between two consecutive antennas and $\lambda$ is the wavelength
corresponding to the central observing frequency $\nu_0$. OWFA has a
high degree of redundancy in baselines. For OWFA, any given baseline
$\mathbf{U}_{a}$ occurs $(264-a)$ times in the array. This can be used
to both calibrate the antenna gains (independent of the sky model) as
well as to estimate the true visibilities (marthi \& Chengalur 2014).

In reality baselines $\mathbf{U}_{a}$ change as frequency
  varies across the observing bandwidth $(B)$. This is an extremely
  important factor that needs to be taken into account in the actual
  data analysis. The expected fractional variation in the baseline
  $\Delta U/U$, about the central frequency $\nu_0$ over the bandwidth
  of observation is $\Delta U / U = B/2\nu_0 \sim 4.5\%$ for $B =30$
  MHz. This is not significant enough to consider in our analysis and
  we have kept the baselines fixed at the value, corresponding to the
  central frequency $\nu_0$. The actual bandwidth may be somewhat
  larger than $B =30$ MHz.

We express the telescope's observing frequency bandwidth as $B = N_c
\Delta \nu_c$ where $N_c$ is the number of the frequency channels and
$\Delta \nu_c$ is the channel-width. For our analysis, We have used
$N_c = 300$ with $\Delta \nu_c = 0.1$ MHz.

The measured Visibilities $\mathcal{V}(\mathbf{U}_a, \nu_n)$ can be
expressed as of sum of the HI signal $\mathcal{S}(\mathbf{U}_a,
\nu_n)$ and the noise $\mathcal{N}(\mathbf{U}_a, \nu_n)$, i.e.,
\begin{equation}
\mathcal{V}(\mathbf{U}_a, \nu_n) = \mathcal{S}(\mathbf{U}_a, \nu_n) +
\mathcal{N}(\mathbf{U}_a, \nu_n)
\end{equation}
assuming that foregrounds have been completely removed from the data.

For the Fisher matrix analysis, it is of convenience to decompose the
visibilities $\mathcal{V}(\mathbf{U}_{a}, \nu_n)$ into delay channels
$\tau_{m}$ (Morales 2005) rather than frequency channels $\nu_n$,
i.e.,
\begin{equation}
v(\mathbf{U}_{a}, \tau_m) = \Delta \nu_c \sum_{n=1}^{N_c} \, e^{2 \pi
  i \tau_{m} \nu_{n}} \, \mathcal{V}(\mathbf{U}_{a}, \nu_n)
\end{equation}
where $\tau_m$ is the delay channel which is defined as follows, i.e.,
\begin{equation}
\tau_m = \frac{m}{B} \hspace{0.5cm} \text{with} \hspace{0.5cm}
\frac{N_c}{2} < m \leq \frac{N_c}{2}
\label{eq:tau}
\end{equation} 

The visibilities $v(\mathbf{U}_{a}, \tau_m)$ and $v(\mathbf{U}_{b},
\tau_n)$ are uncorrelated for $m \neq n$ (Paper II).  It is therefore
necessary to only consider the visibility correlations with $m = n$
for which we define the visibility covariance matrix
\begin{equation}
  C_{ab}(m) = \langle v(\mathbf{U}_{a}, \tau_m) v^{*}(\mathbf{U}_{b},
  \tau_m) \rangle \,.
\end{equation}
The visibility covariance matrix $C_{ab}(m)$ can be expressed in terms
of the redshifted HI 21-cm brightness temperature power spectrum
$P_{T}(\mathbf{k}_{\perp}, k_{\parallel})$ as (eq. (5) of Paper II)
\begin{align}
& C_{ab}(m) = \frac{B}{r_{\nu}^2 r_{\nu}^{'}} \left(\frac{2
    k_B}{\lambda^2}\right)^2 \int d^2 U^{'} \tilde{A}(\u_a-\u^{'})
  \tilde{A}^{*}(\u_b-\u^{'}) \nonumber \\ & \times
  P_{T}(\mathbf{k}_{\perp}, k_{\parallel}) + \frac{2 \Delta \nu_c B
    \sigma_N^{2}}{(264-a)}\delta_{a,b}
\label{eq:cov}
\end{align}
where the first and the second terms refer to the signal and noise
covariance respectively. Here $r_{\nu}$ is the co-moving distance
between the observer and the region of space from where the HI
radiation originated, $r^{'}_{\nu}=\frac{dr_{\nu}}{d \nu}$ gives the
conversion factor from frequency to co-moving distance ($r_{\nu} =
6.85$ Gpc and $r_{\nu}^{'} = 11.5$ Mpc MHz$^{-1}$ for OWFA) and
$\tilde{A}(\mathbf{U})$ is the Fourier transform of the OWFA primary
beam pattern (eq. (6) of Paper I)). The factor $\frac{2
  k_B}{\lambda^2}$ gives the conversion from brightness temperature to
specific intensity (where $k_{B}$ is the Bultzmann constant),
$P_{T}(\mathbf{k_{\perp}}, k_{\parallel})$ is the redshifted HI 21-cm
brightness temperature power spectrum, $\mathbf{k_{\perp}} = \pi
(\mathbf{U}_a + \mathbf{U}_b)/r_{\nu}$ and $k_{\parallel} = 2 \pi
\tau_m/r_{\nu}^{'}$ respectively refer to the perpendicular and
parallel components of the wavevector $\mathbf{k}$ with $k =
\sqrt{k_{\perp}^2 + k_{\parallel}^2}$ (Bharadwaj, Sarkar \&
  Ali 2015). The rms. noise of the measured visibilities has
  the contribution from the system noise and has a value $\sigma_N =
  6.69 \, {\rm Jy}$ for $16 \, {\rm s}$ integration time (Table 1 of
  Paper I) and the factor $(264-a)^{-1}$ in the noise contribution
  accounts for the redundancy in the baseline distribution for OWFA.

For OWFA, the visibilities at any two baselines $\mathbf{U}_a$ and
$\mathbf{U}_b$ are uncorrelated ($C_{ab}(m) = 0$) if $|a-b| > 1$
ie. the visibility at a particular baseline $\mathbf{U}_a$ is only
correlated with the visibilities at the same baselines or the adjacent
baselines $\mathbf{U}_{a \pm 1}$.  Thus, for a fixed $m$, $C_{ab}(m)$
is a symmetric, tridiagonal matrix where the diagonal represents the
visibility correlation at the same baseline whereas the upper and
lower diagonals represent the visibility correlation between the
adjacent baselines. Figure 1 of Paper II shows the signal contribution
for the diagonal and off-diagonal terms of $C_{ab}(m)$.  The
covariance at adjacent baselines is approximately one fourth of the
covariance at the same baselines.  Further, the noise contributes only
to the diagonal terms and it does not figure in the off-diagonal
terms.

We have used the Fisher matrix (eq. (8) of Paper II)
\begin{equation}
F_{\alpha \gamma} = \frac{1}{2} \sum_m C^{-1}_{ab}(m)
[C_{bc}(m)]_{,\alpha} C^{-1}_{cd}(m) [C_{da}(m)]_{,\gamma}
\label{eq:fisher}
\end{equation}
to predict the accuracy with which it will be possible to constrain
the value of various parameters using observations with OWFA.  The
indices $\alpha,\gamma$ here refer to the different parameters whose
values we wish to constrain. The inverse of the Fisher matrix
$F_{\alpha \gamma}$ provides an estimate of the error-covariance
(Dodelson 2003) for these parameters.  In eq. (\ref{eq:fisher}), the
indices $a,b,c,d$ are to be summed over all baselines.  We have used
eq. (\ref{eq:cov}) to calculate the data covariance matrix $C_{ab}(m)$
and also its derivatives $[C_{ab}(m)]_{,\alpha}$ with respect to the
parameters whose values we wish to constrain.  A discussion of the
parameters considered for the present analysis follows in the next
section.

\section{Modelling and Binning the HI power spectrum}
The redshifted HI 21-cm brightness temperature power spectrum
$P_{T}(\mathbf{k})$ is the quantity that will be directly measured by
any cosmological 21-cm experiment. This quantifies the fluctuations in
the brightness temperature originating from two different sources -
1. the intrinsic fluctuations of the HI density in real co-moving
space and 2. the peculiar velocities which introduce brightness
temperature fluctuations through redshift space distortion. For the
purpose of this work, we have modelled $P_{T}(\mathbf{k})$ as
\begin{equation}
 P_{T}(\mathbf{k}) = (1 + \beta \mu^2)^2 P_{T}^r(k)
\end{equation}
where $P_{T}^r(k)$ is the power spectrum of the HI 21-cm brightness
temperature fluctuations in real space and the factor $(1 + \beta \mu^2)^2$
quantifies the the effect of linear redshift space distortion due to
peculiar velocities (Kaiser 1987; Bharadwaj, Nath \& Sethi 2001; Ali
\& Bharadwaj 2014). Here $\beta$ is the redshift space distortion
parameter and $\mu = k_{\parallel}/k$.

OWFA will probe the $k_{\perp}$ and $k_{\parallel}$ - range $1.9
\times 10^{-3} \leq k_{\perp} \leq 5 \times 10^{-1}$ Mpc$^{-1}$ and
$1.8 \times 10^{-2} \leq k_{\parallel} \leq 2.73$ Mpc$^{-1}$
respectively, thereby covering the $k$-range $ 1.82 \times 10^{-2}
\leq k \leq 2.73$ Mpc$^{-1}$. The present work focuses on making
predictions for the accuracy with which it will be possible to measure
$P_{T}^r(k)$ using OWFA. For this purpose, we have divided the entire
$k$-range probed by OWFA into 20 equally spaced logarithmic $k$-bins,
and we use $k^i$ and $[P^r_{T}]^i$ (with $1 \leq i \leq 20)$ to refer
respectively to the average $k$ and $P_{T}^r(k)$ value for each bin.
We have used $\ln ([P_{T}^r]^i)$ and $\ln (\beta)$ as the parameters
for the Fisher matrix analysis (eq. (\ref{eq:fisher})) which gives an
estimate of the precision with which it will be possible to
measure these parameters.

\section{Results and Discussions}
We need a fiducial model for the $P_{T}^r(k)$ and $\beta$ to carry out
the Fisher matrix analysis. We model the $P_{T}^r(k)$ assuming that it
traces the underlying matter power spectrum $P(k)$ with a linear bias
$b_{\rm HI}$ as also assumed by Bharadwaj, Nath \& Sethi (2001),
Bharadwaj \& Sethi (2001), Wyithe \& Loeb (2009),
\begin{equation}
P^r_{T}(k) = (\bar{x}_{\rm HI} \, b_{\rm HI} \, \bar{T})^2 \, P(k)
\label{eq:PHIr}
\end{equation}
where $\bar{x}_{\rm HI}$ is the mean neutral hydrogen fraction. The
characteristic HI brightness temperature $\bar{\rm T}$ is defined as
(Bharadwaj \& Ali 2005)
\begin{equation}
\bar{T}(z) = 4.0\,\text{mK} \, (1+z)^2 \, \left( \frac{1-Y_P}{0.75}
\right) \, \left( \frac{\Omega_b h^2}{0.020} \right) \left(
\frac{0.7}{h} \right) \left( \frac{H_0}{H(z)} \right)
\label{eq:Tbar}
\end{equation} 
where $Y_P$ is the helium mass fraction and the other symbols
  in the above equation have their usual meanings.

The parameter $\beta$ in the observed $P_T(k)$ (eq. \ref{eq:PHIr}) is
defined as $\beta = f(\Omega)/b_{\rm HI}$ where $f(\Omega)$ quantifies
the growth rate of the matter density perturbations, whose value is
specified by the background $\Lambda$CDM cosmological model. Note that
the various terms used in eq. (\ref{eq:PHIr}) correspond to the
redshift, $z = 3.35$ where HI radiation originated.

We have used $\bar{x}_{\rm HI} = 0.02$ for our analysis which
  corresponds to the neutral gas mass density parameter $\Omega_{\rm
    gas} = 10^{-3}$. This value of $\Omega_{\rm gas}$ comes from DLA
  observations in the redshift range of our interest (Prochaska \&
Wolfe 2009; Noterdaeme et al. 2012; Zafar et al. 2013).  Simulations
(Bagla, Khandai \& Datta 2010; Guha Sarkar et al. 2012; Sarkar,
Bharadwaj \& Anathpindika 2016) and analytical modelling (Marin et
al. 2010) suggest the use of a constant, scale-independent bias at
wave numbers $k \leq 1$ Mpc$^{-1}$, and we use the value $b_{HI}=2.0$
for our entire analysis. We use these values and the cosmological
parameters to calculate the fiducial values of $[P_{T}^r]^i$ (pointsin Figure
\ref{fig:err}) and $\beta = 4.93 \times 10^{-1}$.

The value of  $\beta$ can be estimated by sampling the Fourier modes 
$\mathbf{k}$ with a fixed magnitude $k$ which are, however, oriented at
different directions to the line-of-sight. In other words, $\mu =
k_{\parallel}/k$ should uniformly span over the entire range $-1 \leq
\mu \leq 1$. The minimum value of $k_{\parallel}$ probed
  by OWFA is approximately $10$ times larger than the minimum
  value of  $k_{\perp}$. The maximum value of $k_{\parallel}$
  is also $\sim 4$ times larger than  the maximum value of 
  $k_{\perp}$ (Table II of Paper II). In addition to this, the
  sampling width for $k_{\parallel}$ is roughly $\sim 20$ times
  larger than  that of $k_{\perp}$. These disparities  lead to a non-uniform 
  distribution where the  $\k$  modes are largely concentrated around $\mu =
  1$ (see Figure 3 of Paper II). This anisotropic distribution of
  the $\k$ modes does not make OWFA very suitable for measuring $\beta$, 
  and we do not consider this  in our  analysis.

We have considered two different cases for error predictions. In the
first situation, we consider the Conditional errors $\sigma_{ic}$ for
the measurement of the binned HI power spectrum
$[P_{T}^r]^i$. The conditional error $\sigma_{ic}$ represents the
  error on the measurement of $[P_{T}^r]^i$ in a situation 
where the values of all the other
  parameters   are precisely known. Here, we calculate $\sigma_{ic}$ for the 
$i$-th bin by assuming that the values of $\beta$ and $[P_{T}^r]^j$
are precisely known for all the other bins. We use $\sigma_{ic} =
1/\sqrt{F_{ii}}$ to compute the conditional error for the $i$-th bin.

In the second situation, we have considered the Marginalized errors
$\sigma_{im}$ for the measurement of
$[P_{T}^r]^i$. The marginalized error $\sigma_{im}$ gives the
  error on the measurement of $[P_{T}^r]^i$ without assuming any
  prior information about the other parameters. While estimating the
error for the $i$-th bin, we have marginalized over the values of $\beta$ and
$[P_{T}^r]^j$ in the  other bins. In our previous work (Paper II), we have
calculated the marginalized error on the measurement of the amplitude
of the HI power spectrum with a prior on $\beta$ in the range $0.329
\leq \beta \leq 0.986$. In the present work, we have not imposed any
prior on $\beta$ and we have marginalized $\ln ([P_{T}^r]^i)$ and $\ln
(\beta)$ over the entire range $-\infty$ to $+\infty$. We use
$\sigma_{im} = \sqrt{[F^{-1}]_{ii}} $ to calculate the marginalized
error for the $i$-th bin. The conditional  and the marginalized
  errors  here represent the two limiting cases, and  the 
  error estimates would lie somewhere in between  $\sigma_{ic}$ and
  $\sigma_{im}$  if we impose priors on the value of $\beta$  or any of  the
  other parameters.

\begin{figure}
\begin{center}
\psfrag{Conditional}[c][c][0.8][0]{Conditional $\sigma_{ic}$}
\psfrag{Marginalized}[c][c][0.8][0]{Marginalized $\sigma_{im}$}
\psfrag{p3}[c][c][1.0][0]{$\qquad 1,000$ hours}
\psfrag{p2}[c][c][1.2][0]{{Fractional error}}
\psfrag{p1}[c][c][1.2][0]{$k$ Mpc$^{-1}$}
\vskip.2cm \centerline{\hspace{-1.5cm}{\includegraphics[scale
      =.80]{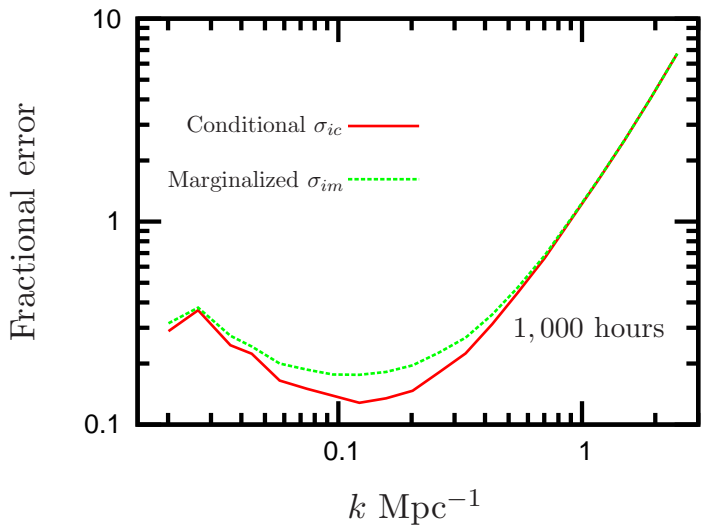}}}
\caption{The figure shows the fractional errors for the
    measurement of $\ln([P_{T}^r]^i)$ in different $k$-bins with
    $1,000$ hours of observation for two situations, conditional
    $\sigma_{ic}$ (red solid line) and marginalized $\sigma_{im}$
    (green dashed line).}
\label{fig:sigma}
\end{center}
\end{figure}

\begin{figure}[h]
\begin{center}
\psfrag{k1}[c][c][1.2][0]{$P_T^r(k)$ Mpc$^{3}$ mK$^{2}$}
\psfrag{k2}[c][c][1.2][0]{$k$ Mpc$^{-1}$} \psfrag{1000
  hr}[c][c][1.0][0]{$1,000$ hours}
\vskip.2cm
\hspace*{-1.5cm}\centerline{{\includegraphics[scale
      =.80]{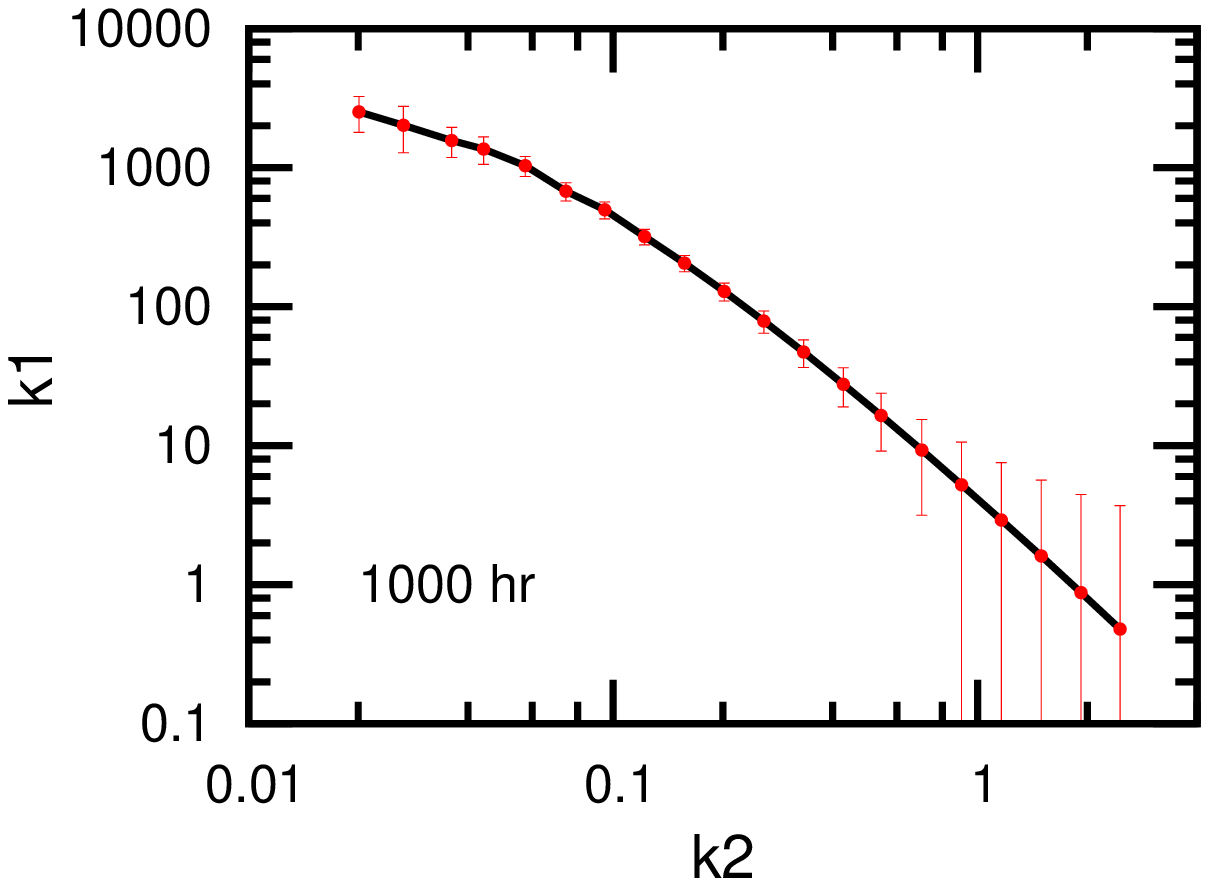}}}
\caption{The binned HI power spectrum $[P_{T}^r]^i$ (points) with
  $1\sigma$ errors (vertical bars) for $1,000$ hours of observation. }
\label{fig:err}
\end{center}
\end{figure}

In Paper II, we have shown that a $5\sigma$ detection of the amplitude
of the $P_{T}^r$ is possible with $\sim 150$ hours of observations.
We therefore need to consider an observing time $t > 150$ hours for
measuring the $[P_{T}^r]^i$ in different $k$-bins. Figure
\ref{fig:sigma} shows both the conditional ($\sigma_{ic}$) and
marginalized ($\sigma_{im}$) errors for $1000$ hours of
observation. Here $\sigma_{ic}$ and $\sigma_{im}$ are respectively the
conditional and marginalized errors for different $\ln([P^r_T]^i)$ which are
the parameters for the Fisher matrix analysis. Here $\sigma_{ic}$ and
$\sigma_{im}$  represent the two 
limiting cases for the error estimates. We expect the error estimates to lie
between these two limiting values in case we impose a prior on the
value of $\beta$ (Paper II). 

We find that the values of $\sigma_{ic}$ and $\sigma_{im}$ agree within
$15\%$, except at the $k$-bins lying in the range $0.06 \leq k \leq 0.3$
Mpc$^{-1}$ where the difference is $\sim 20-35\%$. This
suggests that $\sigma_{ic} = 1/\sqrt{F_{ii}}$ and $\sigma_{im} =
\sqrt{[F^{-1}]_{ii}}$ are not significantly different, indicating that
the contribution from the off-diagonal terms of  $F_{ij}$ are
small. We therefore conclude that the measurements of  $P_{T}^r$ in
different $k$-bins are by and large uncorrelated. In the subsequent
analysis, we have used $\sigma_{ic}$ for predicting errors on the
measurements of $[P_{T}^r]^i$ in different $k$-bins.

Figure \ref{fig:err} shows the binned HI power spectrum $[P_{T}^r]^i$
with the $1\sigma$ errors $\Delta [P_{T}^r]^i = \sigma_{ic} \times
[P_{T}^r]^i$ for $1,000$ hours of observation . The error $\Delta
[P_{T}^r]^i$ on the measurement of the $P_{T}^r$ in a given $k$-bin is
the combination of contributions from the system noise and the cosmic
variance. The noise term in eq.(\ref{eq:cov}) is suppressed by the
factor $(264-a)^{-1}$ due to the redundancy of the OWFA baselines. We
see that the noise contribution goes up as the baseline number $a$ is
increased. The small $k$-bins which correspond to small baselines,
have smaller noise contribution than the larger $k$-bins which
correspond to large baselines. Here we have used logarithmic binning
where the bin width and the number of $k$-modes in a bin increase with
$k$. The cosmic variance in a given $k$-bin goes down with number of
$k$-modes in that bin. We therefore expect the cosmic variance to be
maximum at the smallest $k$-bin and decrease with increasing $k$. As a
whole the errors at smaller $k$-bins are dominated by the cosmic
variance whereas at larger $k$-bins, the errors are dominated by the
system noise.

We can see from Figure \ref{fig:sigma} that $\sigma_{ic}=\Delta
[P_{T}^r]^i/[P_{T}^r]^i$, which is the relative error on the binned
power spectrum, is minimum in the range $k \sim 0.1 - 0.2 \, {\rm
  Mpc}^{-1}$. The cosmic variance dominates the relative error at
smaller values of $k$ $(< 0.1\, {\rm Mpc}^{-1}) $ whereas the system
noise dominates at larger $k$ values $(>0.2 \, {\rm Mpc}^{-1}) $.  We
also see that the relative error is lower than $0.2$ in the range
$0.05 \leq k \leq 0.3 \, {\rm Mpc}^{-1}$ where our results predict a
$5 \sigma$ detection of the binned power spectrum (Figure
\ref{fig:err}).

\begin{figure}
\begin{center}
\psfrag{k2}[c][c][1.5][0]{$t$ hours} \psfrag{k1}[c][c][1.5][0]{$k$
  Mpc$^{-1}$} \psfrag{k3}[c][c][1.5][0]{SNR}
\psfrag{Ten}[c][c][1][0]{$\mathbf{10 \sigma}$}
\psfrag{sev}[c][c][1][0]{$\mathbf{7 \sigma}$}
\psfrag{fiv}[c][c][1][0]{$\mathbf{5 \sigma}$}
\psfrag{thr}[c][c][1][0]{$\mathbf{3 \sigma}$}
\vskip.2cm
\centerline{{\includegraphics[angle=-90,scale=1.0]{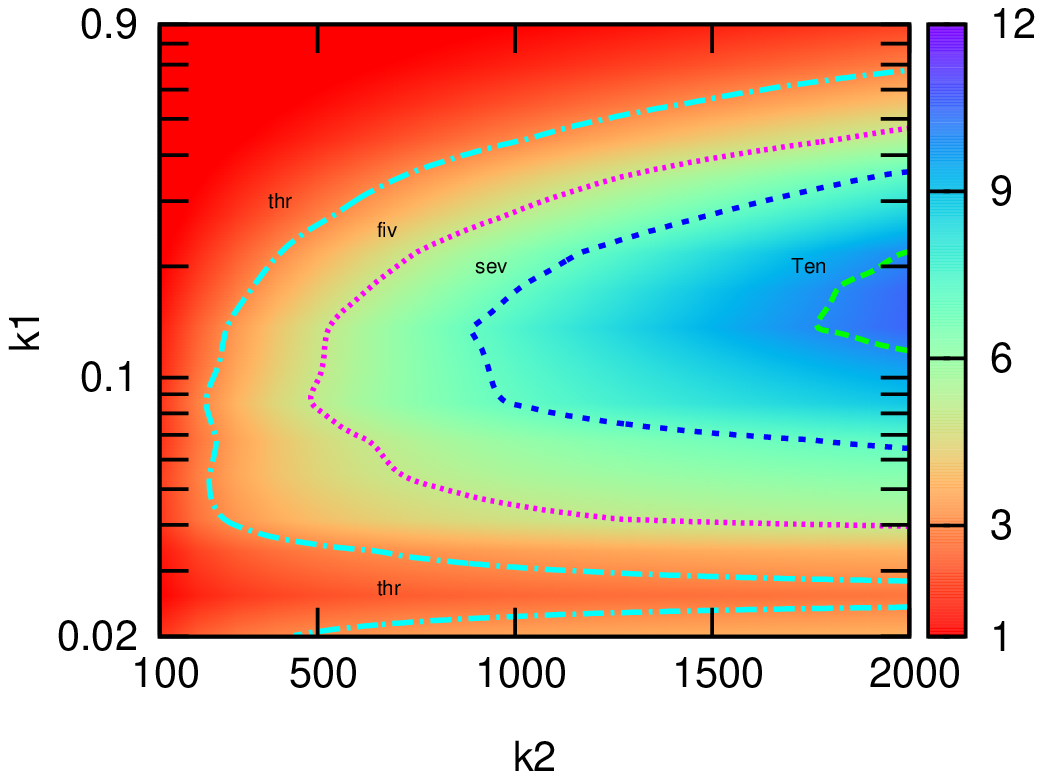}}}
\caption{The SNR contours as function of $k$-bin and observations time
  $t$.}
\label{fig:cont}
\end{center}
\end{figure} 
  
We have so far considered the errors on the measurement of
$[P_{T}^r]^i$ with a given hours of observing time ($1,000$ hours). We
shall now try to understand how the errors $\sigma_{ic}$ vary with
observation time $t$. The time dependence of the visibility covariance
$C_{ab}(m)$ (eq. (\ref{eq:cov})) comes in through the rms. noise of
the measured visibilities $\sigma_N$ which scales inversely with
$\sqrt{t}$, ie. $\sigma_N \sim 1/\sqrt{t}$. We expect the visibility
covariance $C_{ab}(m)$ to vary inversely with $t$, ie. $C_{ab}(m) \sim
1/t$ for small observing times where the noise contribution is
considerably larger than the signal, and we expect $C_{ab}(m)$ to have
a constant value, independent of the observing time, for large values
of $t$.  The derivatives of the $C_{ab}(m)$ which appear in the Fisher
Matrix (eq. (\ref{eq:fisher})) are independent of $t$. It then follows
that the Fisher matrix $F_{\alpha \gamma}$ scales as $F_{\alpha
  \gamma} \sim t^2$ for small observation times and $F_{\alpha
  \gamma}$ has a constant value for large $t$.  We therefore expect
the relative errors $\sigma_{ic}$ to vary as $\sigma_{ic} \sim 1/t$
for small observing times, and become independent of $t$ for large
observation times where the error is dominated by the cosmic variance.

Figure \ref{fig:cont} shows a contour plot of the Signal-to-Noise
Ratio (SNR)
\begin{equation}
\text{SNR} = \frac{1}{\sigma_{ic}} = \frac{[P_{T}^r]^i}{\Delta
  [P_{T}^r]^i},
\end{equation}
as functions of the Fourier mode $k$ and observation time $t$.  We see
that a statistically significant measurement ( $3\sigma$) of the
binned power spectrum is only possible for observations times greater
than $200 \, {\rm hours}$.  A $3\sigma$ detection of $[P_{T}^r]^i$ is
possible in the $k$-range $0.04 \leq k \leq 0.2$ Mpc$^{-1}$ with $200
- 300$ hours of observation. Detection at a significance of $5\sigma$
is not possible with $t \leq 500$ hours of observation. We find that a
$5\sigma$ detection of $[P_{T}^r]^i$ is possible for $0.05 \leq k \leq
0.3$ Mpc$^{-1}$ with $1,000$ hours of observing time. Note that the
SNR peaks in the range $k \sim 0.1 - 0.2$ Mpc$^{-1}$.  In this $k$
range the SNR continues to increase with $t$ for the entire $t$ range
shown here and a $10 \sigma$ detection is possible with $2,000$ hours
of observation. At $k < 0.1 \, {\rm Mpc}^{-1}$, the SNR stops
increasing with $t$ beyond a certain point.  The SNR here becomes
dominated by the cosmic variance as $t$ is increased, and the SNR
contour becomes parallel to the $t$ axis. We see that
  irrespective of the observing time, a $5 \sigma$ detection is not
  possible for $k < 0.036 \, {\rm Mpc}^{-1}$ if only one pointing is
  considered. For $k > 0.2 \, {\rm Mpc}^{-1}$ the error is system
noise dominated, and the SNR continues to increase with increasing
$t$. However, we see that a $5 \sigma$ detection is not possible for
$k > 0.5 \, {\rm Mpc}^{-1}$ within $2,000$ hours of observation.

\begin{figure}[h]
\begin{center}
\psfrag{p2}[c][c][1.2][0]{SNR} \psfrag{p1}[c][c][1.2][0]{$t$ hours}
\psfrag{k1}[c][c][0.9][0]{\hspace*{-1.5cm}$0.036 \, {\rm Mpc}^{-1}$}
\psfrag{k2}[c][c][0.9][0]{\hspace*{-1.5cm}$0.15 \, {\rm Mpc}^{-1}$}
\psfrag{k3}[c][c][0.9][0]{\hspace*{-1.5cm}$0.33 \, {\rm Mpc}^{-1}$}
\psfrag{k4}[c][c][0.9][0]{\hspace*{-1.5cm}$1.16 \, {\rm Mpc}^{-1}$}
\vskip.2cm \centerline{{\includegraphics[scale=1.0]{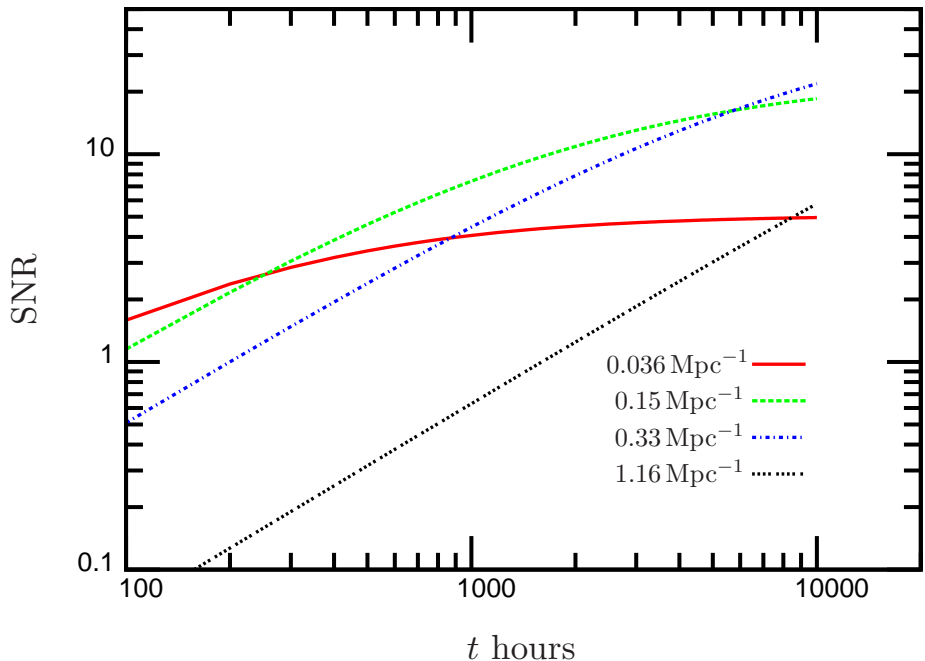}}}
\caption{The Signal-to-Noise Ratio (SNR) with observation time $t$ for
  different $k^i$ values (mentioned in the figure).}
\label{fig:snr}
\end{center}
\end{figure}

As mentioned earlier, we expect SNR $\propto\, t$ for small observing
times when the error is system noise dominated, and we expect the SNR
to saturate at a fixed value for large observing times where the
cosmic variance dominates.  Figure \ref{fig:snr} shows how the SNR
changes with observing time $t$ for a few representative $k$-bins. The
small $k$ bins have a relatively large cosmic variance.  We see that
the SNR at the smallest $k$ bin ($0.036 \, {\rm Mpc}^{-1}$) shown in
this figure is nearly saturated at a very small observing time ($t
\sim 300$ hours), and increases very slowly for larger observing
times.  A $5 \sigma$ detection in this bin requires $\sim 10,000$
hours of observation.  The $k$-bin at $0.33 \, {\rm Mpc}^{-1}$ shows
the ${\rm SNR} \propto t$ scaling for $t \le 700 \, {\rm hours}$,
beyond which the increase in SNR is slower. The two larger $k$-bins
shown in the figure show the ${\rm SNR} \propto t$ behaviour over the
entire $t$ range considered here. However, note that the largest
$k$-bin with $k^i = 1.16 \, {\rm Mpc}^{-1}$ shown in the figure has a rather low
SNR, and a $5 \sigma$ detection is only possible with $10,000 \, {\rm
  hours}$ of observation. 

\begin{figure}[h]
\begin{center}
\psfrag{p2}[c][c][1.2][0]{$t_{\rm CV}$ hours}
\psfrag{p1}[c][c][1.2][0]{$k \, {\rm Mpc}^{-1}$}
\vskip.2cm \centerline{{\includegraphics[scale=1.0]{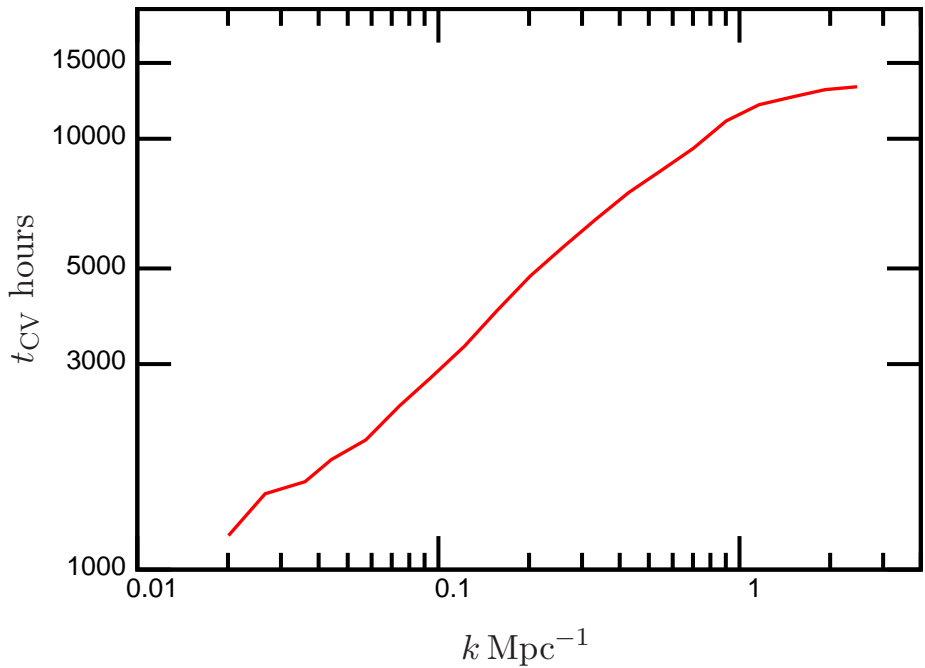}}}
\caption{The observation time ($t_{\rm CV}$) at which the cosmic
  variance starts to dominate the error on the measurement of the HI
  power spectrum for different $k$-bins.}
\label{fig:snrsp}
\end{center}
\end{figure}

As mentioned earlier, we expect the SNR to increase with time
  $t$, ie. ${\rm SNR} \propto t$ for small observing times. The
  increase in the SNR slows down for larger observing times, and for
  even larger observing times $(t \ge t_{\rm CV})$ the SNR saturates
  at a fixed value which is determined by the cosmic variance in the
  particular $k$-bin.  Here $t_{\rm CV}$ referes to the observing time
  beyond which the SNR is determined by the cosmic variance, and we
  have estimated this for the different $k$ bins such that the SNR
  increases slower than $t^{0.002}$ for $(t \ge t_{\rm CV})$. Figure
  \ref{fig:snrsp} shows $t_{\rm CV}$ for the different $k$-bins. For
  any particular bin, it is not possible to increase the SNR any
  further by increasing the observing time beyond $t_{\rm CV}$. We
  find that $t_{\rm CV}$ increases approximately as $t_{\rm CV}
  \propto k^{0.63}$ for $k \leq 1.0 \, {\rm Mpc}^{-1}$. The increase
  in the $t_{\rm CV}$ is rather slow for $k \geq 1 \, {\rm Mpc}^{-1}$
  and saturates at $t_{\rm CV} \sim 15000 \, {\rm hr}$ beyond $k = 2.0
  \, {\rm Mpc}^{-1}$. This behaviour is decided by a combination of
  several factors including the OWFA baseline redundancy, the sampling
  of the 3D Fourier modes and the logarithmic binning.

The discussion till now has entirely focused on observations
  in a single field of view. As already mentioned, the ${\rm SNR}$
  ceases to increase with observing time once $t \sim t_{\rm CV}$. We
  see that $t_{\rm CV} \sim 1,000 \, {\rm hr}$ at the smallest $k$
  bin. The ${\rm SNR}$ in this bin will saturate for $t > 1,000 \,
  {\rm hr}$ and it is necessary to observe multiple fields of view to
  increase the ${\rm SNR}$ any further. The ${\rm SNR}$ scales as
  ${\rm SNR} \propto \sqrt{N}$, where $N$ is the number of fields of
  view.  A possible observational strategy for OWFA would be to
  observe multiple fields of view, with each field being observed for
  a duration of $1,000 - 2,000 \,{\rm hr}$. The $3 \sigma$ contour in
  Figure \ref{fig:cont} would correspond to $5.2 \sigma$ for
  observations in $N=3$ fields of view.  We see that a $5 \sigma$
  detection is possible in nearly all the $k$ bins if $3$ fields of
  view are observed for $1,500 \, {\rm hr}$ each.

\section{Summary and Conclusions}
We have considered Phase II of OWFA to study the prospects of
measuring the redshifted 21-cm power spectrum in different $k$-bins.
The entire analysis is restricted to observations in a single field of
view.  We find that a $5 \sigma$ detection of the binned power
spectrum is possible in the $k$ range $0.05 \leq k \leq 0.3 \, {\rm
  Mpc}^{-1}$ with $1,000$ hours of observation.  The ${\rm SNR}$ peaks
in the $k$ range $0.1 - 0.2 \, {\rm Mpc}^{-1}$ where a $10 \sigma$
detection is possible in $2,000$ hours of observation.  Our study
reveals that it is not very advantageous to observe much beyond
$1,000$ hours as the error in measuring the power spectrum become
cosmic variance dominated in several of the small $k$-bins, and the
${\rm SNR}$ in these bins increase rather slowly with increasing $t$.

As discussed earlier, the variation of the baseline over the observing
bandwidth is $\sim 5\%$. This makes both the diagonal and off-diagonal
components of the Fisher matrix to change, which is expected to be not
more than $5-10\%$.

The redshifted 21-cm signal provides an unique way to measure the BAO
in the post-reioinization era ($z \leq 6$).  This is perceived to be a
sensitive probe of the dark energy. The BAO is a relatively small
feature ($\sim 10 - 15 \%$) that sits on the HI power spectrum. The
five successive peaks of the BAO span the $k$ range $0.045 \leq k \leq
0.3$ Mpc$^{-1}$, which is well within the $k$-range probed by OWFA.
The detection of the BAO requires measuring
the HI power spectrum at a significance of $50\sigma$ or more. From
Figure \ref{fig:cont}, we find that such a sensitivity cannot be
achieved in the relevant $k$ range within $t \sim 2000$ hours of
observation. It is also clear that the required sensitivity cannot be
achieved by considering observations in a few fields of view.  For
detecting the BAO it is necessary to consider a different
observational strategy covering the entire sky (e.g. Shaw et
al. 2013). We plan to address this in future work.

\section*{Acknowledgement}
The authors acknowledge Jayaram N. Chengalur, Jasjeet S. Bagla,
Tirthankar Roy Choudhury, C.R. Subrahmanya, P.K. Manoharan and
Visweshwar Ram Marthi for helpful discussions. AKS would like to
acknowledge Rajesh Mondal and Suman Chatterjee for their help. AKS
would like to specially mention Debanjan Sarkar for a detailed reading
of the manuscript and providing important suggestions. AKS would also
like to express thanks to the anonymous referees whose critical
suggestions helped towards a major improvement of the language and
presentation of the manuscript. SSA would like to acknowledge CTS, IIT
Kharagpur for the use of its facilities and thank the authorities of
IUCAA, Pune, India for providing the Visiting Associateship programme.

\end{document}